\documentclass[preprint,12pt]{elsarticle}

\usepackage{amsmath,amsfonts,amssymb}
\usepackage{graphicx}
\usepackage{pbox}
\usepackage{lineno}
\usepackage{soul}
\usepackage{textcomp}
\usepackage{siunitx}
\usepackage{setspace}

\newcommand{\tmu}{\textmu}

\journal{Thin Solid Films}

\begin{document}

\begin{frontmatter}


\title{Metasurfaces for the Infrared Spectral Range Fabricated Using Two-Photon Polymerization}

\author{Micheal McLamb$^{a,}$\footnote[1]{Corresponding Author: mmclam10@uncc.edu tel. 1-704-687-8020}}
\author{Yanzeng Li$^b$}
\author{Paige Stinson$^a$}
\author{Tino Hofmann$^{a}$}
\address{$^a$Department of Physics and Optical Science, University of North Carolina at Charlotte, Charlotte, NC, USA}
\address{$^b$Department of Chemistry, University of Chicago, Chicago, 5735 S Ellis Ave, Chicago, IL 60637, USA}

\begin{abstract}
\doublespacing
Fabrication of metasurfaces is often time consuming and expensive, involving complex lithographic processes. The maskless fabrication of metasurfaces composed of rectangular Au bars is reported as a suitable alternative, providing cost-eﬀective, rapid prototyping of metasurfaces. The investigated metasurfaces were fabricated using a simple three-step process which is discussed in detail. The fabrication process establishes a simple method for producing high fidelity 2D patterns suitable to synthesize metasurfaces for chemical sensing, beam steering, and perfect reflection/transmission. Comprehensive polarization-sensitive reflection data reveal multiple resonances in the infrared spectral range. In addition to the dipole and substrate resonances, a resonance which is attributed to a coupling between the excitation of the metasurface and the substrate phonon mode is observed.
\end{abstract}

\end{frontmatter}
\doublespacing
\section{Introduction}
Metamaterials describe a group of materials designed to have properties typically not found or existing in naturally occurring compounds \cite{WalserCMIBLID_2001}. The optical properties of metamaterials can be tuned by the shape, geometric parameters, composition, and the arrangement of their constituents \cite{sarychev2007electrodynamics, ZhangAS5_2018_PhaseChange, PengOC412_2018_PolConversion, WanP12_2017_PITTheory, Hering10544_2018_DLWforCalibration, AlipourISJ18_2018_TunableTransparency}.

Metasurfaces are 2D metamaterials which have drawn substantial attention in recent years due to their ability to provide large local electromagnetic field improvements for directivity, surface-enhanced Raman scattering, and vibrational mode coupling, for instance \cite{WangMt21_2018, KimO3_2016_ENZforantenna, BraunAS1_2016_ACS_DLWBarReson}. Plasmonic metasurfaces are a subset of metasurfaces for which a coupling with localized surface plasmon excitations further improves the electromagnetic characteristics of the metasurface \cite{WangMt21_2018}.

 \begin{figure*}
  \centering
   \includegraphics[width=0.75\linewidth, keepaspectratio=true, trim=0 -30 0 0, clip]{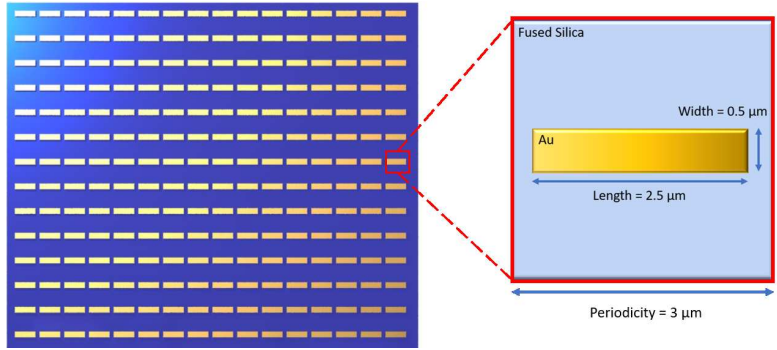}
    \caption{The metasurface geometry studied here is composed of Au bar resonators positioned in a square unit cell on a silica substrate. The inset shows a magnified view of the unit cell. The optimal geometric parameters for a dipole resonance centered at 6~\tmu m were found using finite element model simulations (length = 2.5~\tmu m, width = 0.5~\tmu m, and periodicity = 3~\tmu m). The thickness of the Au bar resonators is 50~nm.}
    \label{fig:unitcell}
\end{figure*}

Metasurfaces have been demonstrated using a wide range of constituents from nanorods, nanodots, to nanobars for which high sensitivity photodetection, hot electron collection, and biosensing has been observed \cite{wu2016infrared}. There are also promising applications in energy harvesting, thermal imaging, and thermal emission \cite{wu2016infrared,hobbs2016optimized, Julianapa_2019_ThermalImaging}. Au nanobars have recently been used in plasmonic sensing where they are superior to traditional structures by offering advantages in parameter control through changes in structure and index \cite{wu2016infrared}.

Several different fabrication methods ranging from photolithography/E-beam deposition/etching/lift-off to laser induced forward transfer approaches, have been explored for the fabrication of metasurfaces \cite{SuOe26_2018_AdvinFabandApp, GenevetO4_2017_RecentAdvs, WangMt21_2018}. In addition, significant progress has been made in enhancing the fabrication scale of these metasurfaces using self-assembly processes \cite{SuOe26_2018_AdvinFabandApp}. While self-assembly processes, in principle, allow scaling to larger metasurface areas, the accessible constituent geometries are often limited. 

Recently, direct laser writing using two-photon polymerization has been explored for the fabrication of metasurfaces \cite{SuOe26_2018_AdvinFabandApp, GenevetO4_2017_RecentAdvs, BraunAS1_2016_ACS_DLWBarReson}. This approach has the advantage that it does not require the use of expensive lithographic masks and therefore reduces fabrication time and cost. It further enables the rapid prototyping of metasurfaces with different constituent geometries.

In this study we report on the fabrication and polarization optical characterization of metasurfaces in the infrared spectral range. The fabrication concept presented here further simplifies the two-photon polymerization configuration and improves the quality of the fabricated metasurfaces compared to previously used approaches \cite{BraunAS1_2016_ACS_DLWBarReson}.
\section{Design and Fabrication}

\subsection{Modeling}
Finite element model simulations were performed using COMSOL in order to determine the optimal geometric parameters of a metasurface composed of bar resonators to produce a strong dipole resonance in the infrared spectral range. The calculations where carried out using periodic boundary conditions with a unit cell as shown in the inset of Fig.~\ref{fig:unitcell}. The unit cell consists of a single rectangular Au bar resonator, the fused silica substrate, and ambient air. Accurate dielectric function data were obtained for the fused silica substrate (Nanoscribe GmbH) using ellipsometric measurements in the range from 1.4~\tmu m - 33~\tmu m (IR-Vase, J.A. Woollam Co., Inc.) and imported into COMSOL. The dielectric function for Au is based on a two-term Drude model imported from optical material library of WVase (J.A. Woollam Co., Inc.) into COMSOL without any further variation. 
 
The width and length of the rectangular dipole as well as the periodicity of the array were varied in order to determine dimensions that would yield a resonance centered at 6~\tmu m. The height of the resonators was held at a constant 50~nm. As a result, the bar resonator length and width were determined to be 2.5~\tmu m and 0.5~\tmu m, respectively. 
 
\subsection{Two-photon polymerization-based nano-patterning}
\begin{figure}
   \centering
   \includegraphics[width=0.8\linewidth, keepaspectratio=true, trim=45 -20 0 140, clip]{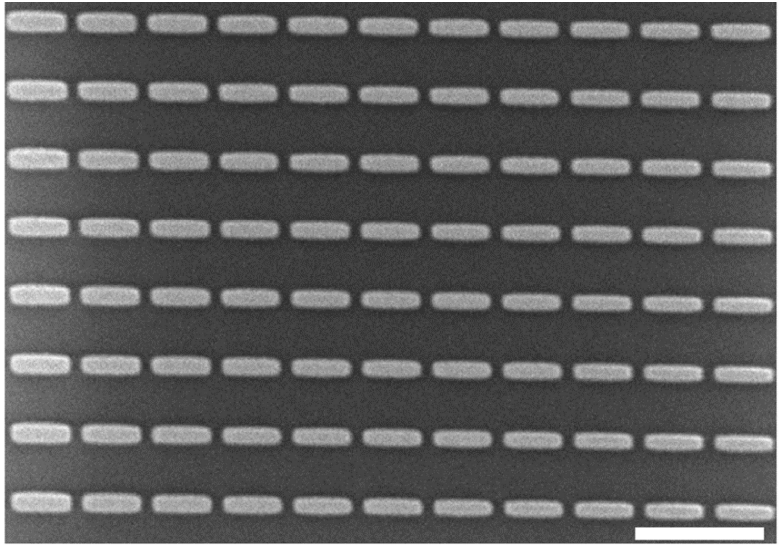}
    \caption{Top-view SEM micrograph of the metasurface depicting the regular arrangement of the rectangular bar resonators. The scale bar at the bottom right indicates 5~\tmu m. The bar resonator has a long axis length of 2.3~\tmu m and a short axis length of 0.5~\tmu m. }
    \label{fig:sem}
\end{figure}

The investigated metasurfaces were fabricated using a three-step process onto fused silica substrates [(25 $\times$ 25)~mm$^{2}$, 700~\tmu m thick, Nanoscribe GmbH] which were cleaned prior to the deposition by rinsing with isopropanol-2 and drying with N$_2$. 
First, a patterned sacriﬁcial IP-Dip layer was deposited using a two-photon polymerization process, which allows its maskless fabrication. The sacrificial layer has a thickness of 200~nm and is inversely patterned to the metasurface depicted in Fig.~\ref{fig:unitcell}. This step is followed by metallization and subsequent sacriﬁcial layer lift oﬀ. As a result, a metasurface with high pattern ﬁdelity is obtained as shown in Fig.~\ref{fig:sem}.    

A commercial two-photon lithography system (Photonic Professional GT, Nanoscribe GmbH) was employed for the patterning of the sacrificial layer. The instrument utilizes a 780~nm femtosecond laser in conjunction with an inverted microscope and a 63x immersion objective. High resolution direct laser writing is achieved by inducing the absorption of two photons \cite{Kim10894_2019_IntricateDLW}. This non-linear process results in the polymerization of a spherical volume with a diameter of approximately 200~nm\cite{Li_2019_DLW, BraunAS1_2016_ACS_DLWBarReson, Kim10894_2019_IntricateDLW}. 

The position of the voxel in the x-y plane is determined by the motorized sample stage while the objective is fixed. Further control of the voxel position is achieved by manipulating the beam using a galvanometer. This allows the position of the voxel to be adjusted within a circular area in the x-y plane with a diameter of approximately 200~\tmu m. For the fabrication of the patterned sacrificial layer the vertical voxel position was at the interface of the substrate and the IP-Dip monomer with no additional changes in the vertical voxel position made. To ensure highest pattern fidelity and interface adhesion the exposure parameters were optimized prior to the sacrificial layer fabrication by adjusting laser power and write speed \cite{li2019near_NSDose, li2018broadband_OL}. 

After polymerization, excess IP-Dip monomer is removed from the substrate by vertically immersing the sample into PGMEA for two minutes and then into isopropanol-2 for two minutes. Once the excess monomer has been removed the sample is allowed to air dry at room temperature for approximately 5 minutes.

\subsection{Metallization and Lift-off}

After the direct laser writing of the patterned sacrificial layer is completed, the sample was metallized (Kurt J.~Lesker PVD 75). Initially, electron beam evaporation was used to deposit a 7~nm adhesion layer of chromium, sourced from Chromium pieces (99.95\%) and, immediately following, a 50~nm Au layer evaporated from Au pellets (99.99\%). Both films were deposited at a rate of 1~$\si{\angstrom}/$s. All depositions were carried out at room temperature at a pressure of 1.9$\times$10$^{-7}$~Torr. The substrate holder was rotated during the deposition at approximately 15 rpm to ensure a homogeneous film thickness across the entire sample.  

After the metallization, the sacrificial photoresist layer was removed by plasma cleaning at 150~watts and a 12~SCCM flow of O$_{2}$ (Tergeo Plus, Pie Scientific). After the plasma cleaning the substrate is completely immersed in acetone until the sacrificial photoresist layer is lifted off completely. Subsequently, the sample is cleaned by rinsing with isopropanol-2 and methanol and dried with N$_2$. 

\section{Results \& Discussion}

The metasurface was characterized using polarized infrared reflection measurements in order to determine its infrared optical response. The experimental infrared reflection data are compared with finite element model based calculated reflection data. Complementary scanning electron microscope images (JEOL 6460LV) were obtained in order to verify the dimensions and integrity of the fabricated metasurface. 

\begin{figure}[hbt]
  \centering
\includegraphics[width=0.9\linewidth, keepaspectratio=true]{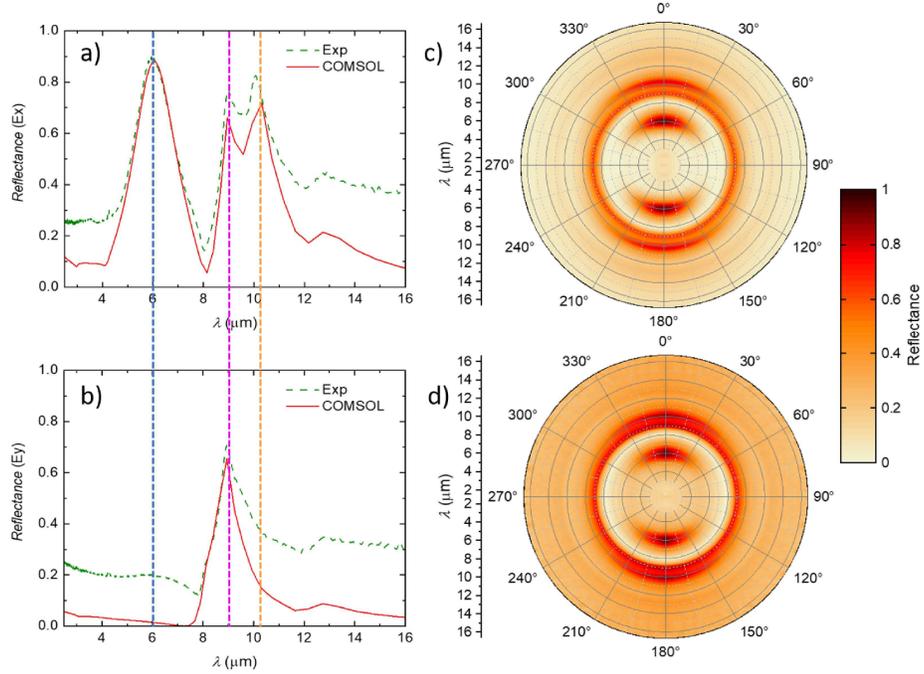}
\caption{The measured (green dashed) and experimental (red solid) reflectance data for,  linear polarization parallel to the long and short axis, is shown in panel (a) and (b), respectively. The dashed vertical lines from left to right indicate the position of the dipole resonance (blue), phonon resonance of the substrate (magenta), and the coupled dipole-phonon resonance (orange). Panels (c) and (d) depict a polar map of the simulated and experimental reflectance, respectively. The incident radiation is linearly polarized along the long axis of the bar resonator when the polarizer is oriented at 0$^{\circ}$.}
    \label{fig:data}
\end{figure}

Fig.~\ref{fig:sem} depicts a SEM micrograph top-view of the fabricated  metasurface. As can be clearly observed, the rectangular bar resonators were fabricated with high fidelity. In particular the true-to-design geometry the rectangular resonator bars is notable if compared to similar structures obtained with different lithographic techniques \cite{SuOe26_2018_AdvinFabandApp, BraunAS1_2016_ACS_DLWBarReson}. The dimensions of the fabricated bar resonators are found to be in good agreement with the nominal values shown in Fig.~\ref{fig:unitcell}. Analysis of the SEM micrograph reveals a length of the long and short axis of the bar resonators of 2.3~\tmu m and 0.5~\tmu m, respectively, which is in good agreement with the nominal design values of the long and short axes. Furthermore, the dimensions of the bar resonators are homogeneous across the metasurfaces. 

Linearly polarized reflectance data was captured using a Hyperion 3000 microscope (Bruker) in combination with a Vertex 70 FTIR spectrometer (Bruker). The reflectance measurements where carried out in the spectral range of 1.3~\tmu m to 16~\tmu m with a resolution of 0.05~\tmu m. 
A 15x IR Schwarzschild objective and a mercury cadmium telluride detector were used for all measurements. An adjustable, KRS-5, wire-grid polarizer mounted in a custom-built, electro-mechanical, rotation stage was used to polarize the incident radiation. 

The experimental (green dashed lines) and finite element model calculated (red solid lines) reflectance spectra in Fig.~\ref{fig:data} panels (a) and (b) display the response of the bar resonators to linearly polarized light along the long and short axes, respectively. Overall, a good agreement between the model calculated and the experimental data can be observed. All the major spectral features and their relative magnitudes are reproduced by the model calculation. An offset can be noticed at the lower and upper end of the spectral range. 

When the light is polarized along the long axis three distinctive peaks can be observed. The first peak, located at 6~\tmu m, is due to the dipole resonance of the bar resonator. This resonance is polarization sensitive as can be seen by comparing the reflectance spectra in Fig.~\ref{fig:data} panels (a) and (b). While the resonance is dominating the reflectance spectrum obtained with the input polarization oriented parallel to the long axis of the resonator [Fig.~\ref{fig:data} (a)], it is completely suppressed for the perpendicular orientation [Fig.~\ref{fig:data} (a)]. 
This behavior is expected for the dipole resonance associated with a simple bar resonator geometry as shown in Fig.~\ref{fig:unitcell}. This is in good agreement with linearly polarized reflectance data reported by Braun and Maier \cite{BraunAS1_2016_ACS_DLWBarReson}. A similar behavior can be observed for a resonance located at a wavelength of approximately 10.2~\tmu m. For the resonance located at 9~\tmu m the polarization sensitivity is lacking. This resonance is attributed to a phonon excitation in the silica substrate. 

Panels (c) and (d) of Fig.~\ref{fig:data} show the finite element model calculated and experimental polar maps of the polarized reflectance spectra, respectively. The polarization of the incident radiation is varied from 0$^{\circ}$ to 360$^{\circ}$ with a resolution of 1$^{\circ}$. The azimuthal orientation of the polarization is denoted with respect to the long axis of the bar resonators. A polarization of 0$^{\circ}$ and 90$^{\circ}$ therefore corresponds to a polarization state parallel and perpendicular to the long axis of the resonator, respectively. 

At a first glance a good agreement between the experimental and the model calculated data can be recognized. The most notable feature is the dipole resonance, which can be clearly observed at 6~\tmu m with peaks in reflectivity at 0$^{\circ}$ and 180$^{\circ}$ in Fig.~\ref{fig:data} (c) and (d). The polarization behavior of the second polarization dependent resonance at 10.2~\tmu m is also well reproduced. We attribute this resonance to a coupling between the phonon resonance and dipole resonance of the metasurface. Such coupling phenomena have recently attracted much attention, but are outside of the scope of this work. The interested reader is referred to Ref.~\cite{KimO3_2016_ENZforantenna, pollard2009optical, jun2013epsilon} and references therein. 
At 9~\tmu m the polarization independent phonon resonance of the silica substrate can be observed in both panels.    

\section{Conclusion}

We have demonstrated the fabrication of metasurfaces for the infrared spectral range using a three-step process which includes the maskless deposition of a patterned sacrificial layer by two-photon polymerization. A simple metasurface composed of bar resonators arranged on the surface in a square unit cell was used to demonstrate the capabilities of this synthesis approach. The fabricated metasurfaces exhibit features with as-designed dimensions and high structural fidelity.   

A comprehensive infrared-optical characterization using polarization-sensitive reflection measurements was carried out. We found a good agreement between the experimental reflection data and the model-based calculated reflection data. Several reflectance maxima were identified in the spectral range from 1.3~\tmu m to 16~\tmu m. 
One resonance with a distinctive polarization characteristic was attributed to the dipole resonance of the bar resonator. In addition, we identified a resonance in the vicinity of the polarization independent substrate phonon-resonance which exhibits a similar polarization dependence as the dipole resonance. Such coupled excitations between dipole and substrate phonon resonances have been observed for several substrate materials in the infrared spectral range, but so far have not been presented for fused silica. 

In summary, we have shown that direct laser writing allows the maskless fabrication of 2D metasurfaces with high structural fidelity. State-of-the-art direct laser writing systems provide voxel sizes which enable the deposition of patterned sacrificial layers with geometries optimized for the infrared spectral range. 
The maskless, direct laser writing-based, process presented here therefore 
provides a suitable alternative for cost-effective, rapid prototyping of metasurfaces. In contrast to classical fabrication of metasurfaces, the complex and time consuming fabrication of lithographic masks is completely omitted.
We envision that this technique will provide a suitable fabrication approach for rapid prototyping and cost-effective manufacturing of infrared plasmonic metasurfaces with complex geometries and applications in areas such as perfect absorption and index sensing. 

\section*{Acknowledgment}
MM, PS, YL, and TH would like to acknowledge the valuable discussions with Susanne Lee within the NSF IUCRC for Metamaterials. The authors are grateful for support from the National Science Foundation (1624572) within the IUCRC Center for Metamaterials and through the NSF MRI 1828430, the Army Research Office (W911NF-14-1-0299) and the Department of Physics and Optical Science of the University of North Carolina at Charlotte.


\end{document}